%Paper: hep-ph/9503402
%From: Robert Leigh <leigh@physics.rutgers.edu>
%Date: Tue, 21 Mar 1995 18:26:14 -0500 (EST)

\input harvmac
\let\norefs=n
\def\MTitle#1#2#3#4{\nopagenumbers\abstractfont\hsize=\hstitle\rightline{#1}%
\vskip 1in\centerline{\titlefont #2}\vskip .08in\centerline{\titlefont #3}
\vskip .1in\centerline{\titlefont #4}\abstractfont\vskip .5in\pageno=0}
\def\qslsh{/\kern-.5em q}
\def\parslsh{/\kern-.5em \partial}
\def\gsim{{~\raise.15em\hbox{$>$}\kern-.85em
          \lower.35em\hbox{$\sim$}~}}
\def\lsim{{~\raise.15em\hbox{$<$}\kern-.85em
          \lower.35em\hbox{$\sim$}~}}

\def\ie{{\it i.e.,}}
\def\mpl{m_{Pl}}

\MTitle{hep-ph/9503402, RU-94-100}
{Supersymmetry, Finite Temperature}{and Gravitino Production}
{in the Early Universe}
\bigskip
\centerline{Robert G. Leigh and Riccardo Rattazzi}
\smallskip
\centerline{\it Department of Physics and Astronomy}
\centerline{\it Rutgers University}
\centerline{\it Piscataway, NJ 08855-0849, USA}
\bigskip
\noindent
We reconsider post-inflation gravitino production, in the context
of hidden sector supergravity models. We discuss the
possible role of supersymmetry breaking from finite temperature effects
in enhancing the rate for gravitino production and argue that there is
no such enhancement. Our conclusion is based on a simple decoupling argument,
which is independent of temperature. We also characterize the
properties of goldstino-like hydrodynamic fluctuations that arise in
supersymmetric models at finite temperature. We show that they cannot
lead to enhanced gravitino emission via infrared divergences. We
comment on an analogy with axion emission from hadronic matter.

\Date{2/95}
%\Date{\datestamp}

\lref\dinenel{M. Dine and A.E. Nelson,
{\it Phys. Rev.} {\bf D48} (1993) 1277 (hep-ph/9303230);
Santa Cruz preprint SCIPP-94-21, (hep-ph/9408384).}
\lref\girard{L. Girardello, M.T. Grisaru and P. Salomonson, {\it Nucl.
Phys.} {\bf B178} (1981) 331.}
\lref\vanhove{L. Van Hove, {\it Nucl. Phys.} {\bf B207} (1982) 15.}
\lref\boyan{D. Boyanovsky, {\it Phys. Rev.} {\bf D29} (1984) 743.}
\lref\aoyboy{H. Aoyama and D. Boyanovsky, {\it Phys. Rev.} {\bf D30}
(1984) 1356.}
\lref\aoyama{H. Aoyama, {\it Phys. Lett.} {\bf 171B} (1986) 420;
I. Ojima, {\it Lett. Math. Phys.} {\bf 11} (1986) 73.}
\lref\matsum{H. Matsumoto, {\it et al} {\it Physica} {\bf 15D} (1985) 163.}
\lref\pisarski{R.D. Pisarski, {\it Nucl. Phys.} {\bf B309} (1988) 476.}
\lref\weingrav {S. Weinberg, {\it Phys. Rev. Lett.} {\bf 48} (1982) 1303.}
\lref\pagprim {H. Pagels and J.R. Primack,
{\it Phys. Rev. Lett.} {\bf 48} (1982) 223.}
\lref\ekn {J. Ellis, J.E. Kim and D.V. Nanopoulos,
{\it Phys. Lett.} {\bf 145B} (1984) 181.}
\lref\fisch{W. Fischler, {\it Phys. Lett.} {\bf 332B} (1994) 277,
(hep-th/9404044).}
\lref\weldon{V.V. Klimov, {\it Sov. J. Nucl. Phys.} {\bf 33} (1981) 934;
{\it Sov. Phys. JETP} {\bf 55} (1982) 199;
H.A. Weldon, {\it Phys. Rev.} {\bf D26} (1982) 1394; {\bf D26} (1982) 2789.}
\lref\strocchi{F. Strocchi, {\it Elements of Quantum Mechanics of
Infinite Systems}, World Scientific, Singapore, 1985.}
\lref\weinberg{S. Weinberg, {\it Gravitation and Cosmology}, Wiley,
New York, 1972.}
\lref\gund{R. Gudmundsdottir and P. Salomonson,
{\it Nucl. Phys.} {\bf B285} (1987) 1.}
\lref\Weldon{H.A. Weldon, {\it Phys. Rev.} {\bf D28} (1983) 2007.}
\lref\hardthermloops{R.D. Pisarski,
{\it Phys. Rev. Lett.} {\bf 63} (1989) 1129.}
\lref\leptoprod{E. Braaten and R.D. Pisarski and T.C. Yuan,
{\it Phys. Rev. Lett.} {\bf 64} (1990) 2242.}
\lref\schenk{A. Schenk, {\it Nucl. Phys.} {\bf B363} (1991) 97;
{\it Phys.Rev.} {\bf D47} (1993) 5138.}
\lref\fayet{P. Fayet, {\it Phys. Lett.} {\bf 84B} (1979) 421;
Casalbuoni {\it et al.}, {\it Phys. Lett.} {\bf 215B} (1988) 313.}
\lref\eric{E. Braaten, preprint NUHEP-TH-94-24 (hep-ph/9409434).}

\newsec{Introduction}

Supersymmetry is an attractive theoretical framework as it provides a
solution to the hierarchy problem for fundamental scalar masses.
With dynamical mechanisms for supersymmetry breaking, the gauge
hierarchy problem ($m_W\ll\mpl$) can also be explained.

It is of great interest then to study the effects of supersymmetry on
various experimental and cosmological data. There is of course as yet
no direct evidence for the existence of the supersymmetric partners of
the particles of the Standard Model. We must then rely on studying
indirect effects. There has been much work done in the last two
decades in these regards. In this paper, we will focus on certain
cosmological issues.

In particular, the standard cosmological models predict that the
early Universe was filled with a plasma; the constituents of this
plasma have densities determined by equilibrium thermodynamics.
Since the Bose-Einstein and Fermi-Dirac distributions are different,
the plasma is populated by different amounts of on-shell fermions
and bosons.

In a theory in which supersymmetry plays a role,\foot{Here we mean that
supersymmetry is unbroken at some scale in the zero-temperature
theory.} temperature effects may invalidate various cancellations
implied by the symmetry between fermions and bosons. It is important
then to investigate what role supersymmetry really plays in the early
universe, and to study possible effects of temperature on various
calculations in supersymmetric models.

In all extant models of supersymmetry breaking, the theory consists
of two sectors. In the first sector, supersymmetry is broken by
some mechanism which in general we will not specify.
The observable particles, \ie\ those of the
standard model, are in a second sector. The two sectors are
separate in the sense that they interact only very weakly with one
another. In the standard hidden sector scenarios, this interaction
is provided by non-renormalizable gravitational effects. In models
of low-energy supersymmetry breaking, such as those recently discussed
by Dine and Nelson\dinenel, the communication of the two sectors
is mediated by gauge interactions.
We note that in the latter models,
the gravitino can be extremely light ($m_{3/2}=F/m_{Pl}\gsim
m_W^2/m_{Pl}$), and cosmology places no bounds on its interactions,
as long as it is sufficiently light.\weingrav\
We will thus concentrate in this paper on hidden sector models.

We have been particularly
motivated by the recent work of W. Fischler\fisch\ on post-inflation
gravitino production, in which it is suggested that temperature effects
can greatly enhance the Goldstino component of
gravitino production cross-sections, leading to
very tight constraints. The plan of this paper is as follows.
In the next section, we
consider the equivalence theorem for Goldstinoes and gravitinoes
at high energies/temperatures and give a decoupling argument based
on finite-temperature Ward identities that suggests that Goldstino
amplitudes will in fact not be enhanced. In the
third section, we suggest a calculation that can correctly compute
the total Goldstino production rate, making proper use of real-time
finite-temperature quantities. This calculation avoids any direct
discussion of individual Feynman diagrams; we argue that the total
rate can be computed in (re-summed) perturbation theory for weak
coupling. In the second half of this paper, we will review older
literature on the effects of temperature on supersymmetric theories.
We clarify and extend results on how supersymmetry is broken, and its
physical manifestation as propagating modes in the plasma.

\newsec{Goldstino-Gravitino Equivalence and Decoupling}

In this section we consider the possible relevance of gravitino
production in the primordial plasma after reheating. Since we are
dealing with temperatures $T\gsim m_{3/2}$, the transverse and
longitudinal modes of the gravitino field may play very different
roles. In particular the transverse (spin 3/2) component couples
always with gauge strength $1/m_{Pl}$. Thus the rate of change of its
density $\dot n_{3/2}$ is at most  of order $T^6/m_{Pl}^2$. In conventional
cosmological scenarios, this small rate  gives the usual bounds.
If the gravitino is stable, it must be lighter than a few keV, in order
that it not dominate the energy density.\pagprim\ If it does decay,
in the absence of dilution by inflation, nucleosynthesis will
constrain $m_{3/2}\gsim 10$ TeV.\weingrav\ In inflationary scenarios,
gravitinoes will be diluted, with $n_{3/2}/n_{\gamma}\sim
T_{RH}/\mpl$. However if the reheat
temperature $T_{RH}$ after inflation is too high ($T_{RH}\gsim 10^{9-10}$ GeV),
gravitinoes will re-attain sufficient density as to have problematic
effects on nucleosynthesis,
entropy release after nucleosynthesis, light element
photo-dissociation and distortions in the microwave background.\ekn\

The longitudinal (spin 1/2) mode, however, can
interact more strongly. Indeed the interactions of this mode at $E\gsim
m_{3/2}$, are well described by those of the Goldstino via the
equivalence theorem.\fayet\ At high energies they are proportional to a power
of $(E/m_{Pl}m_{3/2})\sim E/F$, and could perhaps\fisch\ lead to
a large $\dot n_{3/2}\propto T^8/F^2$. The cosmological consequences could
then be rather strong. In what follows, however, we argue that in the
elementary particle models of phenomenological interest the rate
of goldstino production is never so enhanced. Indeed it is always
suppressed by at least one power of $1/M^2$, where $M$ is the scale
of the physics communicating SUSY breaking to conventional matter.
In particular, when SUSY breaking is communicated by gravity, we have
$M\sim\mpl$ and the rate for production of helicity 1/2 gravitinos is
never more important than the rate for helicity 3/2.

The models of interest consist of a hidden and an observable sector.
It is assumed that in the absence of interactions between them the
hidden sector breaks supersymmetry while the observable one does not.
%Our aim is to demonstrate that these
%couplings must vanish in the limit $\mpl\rightarrow\infty$, when
%the scale of supersymmetry breaking $F$ is held fixed. This decoupling
%property is not affected by finite temperature.
The fundamental Lagrangian of the models of interest may thus always be
written in the form:
\eqn\lagr{{\cal L}={\cal L}_{obs}(\phi)+{\cal L}_{hid}(X)+\sum_n\epsilon^n
{\cal L}_n(\phi,X)}
where $\epsilon$ is  a small parameter. In models where the
interaction between the two sectors is induced by some physics at a
scale $M$ we have $\epsilon=1/M$. For instance in conventional
supergravity models $M=m_{Pl}$. The dynamics of ${\cal L}_{hid}$
breaks SUSY at a scale $F=M_s^2$. The SUSY breaking effects in the
observable sector are then $O(F/M)$, and it
must be $F/M\sim M_Z$ for phenomenological reasons.
It is manifest from the form of this Lagrangian
that, in the limit $\epsilon\rightarrow0$, the Goldstone fermion
of spontaneously broken supersymmetry does not couple to the fields
$\phi$. The form of eq. \lagr\ then shows explicitly that, as long as
perturbation theory is valid, also at
finite temperature the rate of Goldstino production from observable
matter is suppressed by at least $\epsilon^2$.
 We note that the Goldstino production cross-section of
Ref.\fisch\ does not satisfy this property. We also note that if we
were interested in the production of Goldstinos from a thermal bath
of hidden sector particles we would indeed get a rate scaling like
$T^8/F^2$. However this is not a situation of cosmological interest.

By performing a local field redefinition, the Goldstino coupling with
matter can be written in terms of the supercurrent.
At lowest order in the Goldstino $\chi$ field we have
\eqn\coupl{{\cal L}_{\chi}={1\over F}\partial_\mu\chi\left[ S^\mu_{\phi}
+S^\mu_{X}+\sum_n\epsilon^n S^\mu_n\right]}
where in an obvious notation $S^\mu_\alpha$ are
the contributions to the supercurrent from the terms in eq. \lagr. The
decoupling of goldstino and observable sector is not explicit in
eq. \coupl. When calculating scattering amplitudes the decoupling
shows up via cancellations among  diagrams which are related
to each other by  supersymmetry. The diagrammatic way of reasoning is
however very dangerous when trying to draw conclusions on finite
temperature processes. Indeed the presence of a thermal bath
breaks\foot{See section 4.} supersymmetry (as well as Lorentz invariance), and
one might then naively expect that the ``effective'' diagrams will
be related to each other up to terms proportional to a power of
$T$. That is, since the static limit of propagators are affected by finite
temperature, it is tempting to say that the mass splittings should
be replaced by temperature dependent factors.
As a result of this, the suppression of processes with goldstinos
and ordinary matter would be $O(T/F)$ rather then $\epsilon\sim 1/M$,
in clear contradiction with our decoupling argument.
However, when using eq. \coupl\ to parametrize the goldstino
couplings, one should note that $S^{\mu}_\phi$ is a conserved current in the
limit $\epsilon \to 0$. This means that the decoupling, in this
language,  is the result of a Ward identity.
Real time correlation functions will obey Ward identities
that are identical to their zero temperature counterparts, except that
the correlation functions should be interpreted as finite temperature
averages. In particular, the current $S_\phi$ is indeed conserved in
the limit $M\rightarrow\infty$, and so any correlation function
involving a Goldstino and observable matter must also vanish in this
limit, irrespective of temperature.
The introduction of ensemble parameters like
temperature (or chemical potential for abelian symmetries),
corresponds to the choice of a particular state, which may not be
invariant under some symmetries of the system. Ward identities,
on the other hand, express properties of the dynamics, \ie\ the
Hamiltonian, and as such do not depend on the particular state or
ensemble average. In other words, the introduction of ensemble
parameters can lead to a Goldstone representation of a symmetry but not
to its explicit breakdown. Then an effective
finite temperature description of real time processes, if any  is possible,
must preserve the form of the Ward identities.
This is reminiscent of what happens with chiral symmetry
in QCD, where the  low energy pion lagrangian realizes the same Ward
identities of the fundamental QCD dynamics.

Now one should note that such decoupling arguments can in general
be violated by the effects of resonant enhancement. This occurs when
a relatively large mixing arises amongst nearly degenerate states,
because of a small energy denominator.
However, at finite temperature, because of damping phenomena, coherence
can be maintained only for finite times $\lsim t_{\rm damp}$,
and the resonant enhancement is limited. Indeed the only modes
that are undamped are those corresponding to  Goldstone singularities
at zero energy and momentum. It is known that finite temperature Ward
identities (even in the absence of spontaneous supersymmetry
breaking) require the existence of a fermionic excitation with
some properties similar to a Goldstino. In the next two sections we will show
that these Goldstino-like hydrodynamic modes cannot lead to enhanced
gravitino production, simply because of the form of
their dispersion relation. In the Appendix, we comment on the
case of axion emission\foot{We thank S. Thomas and W. Fischler for
bringing this example to our attention.}
from thermal pions, where a partial enhancement
does indeed take place due to resonant mixing with a Goldstone mode.

%In the next section, we outline a calculation that would correctly
%compute the production of Goldstinoes, making proper use of
%real-time quantities. This calculation avoids any direct
%discussion of individual diagrams, and rather computes the
%complete rate of production.

\newsec{Goldstino production at finite temperature: a calculation}

We now present the outline of a calculation of the production rate of
gravitinoes at finite temperature. It is well known that the interpretation of
finite temperature Green's functions as corresponding to scattering amplitudes
is tricky. Essentially the problem is one of boundary conditions: we do not
have simply asymptotic states to deal with, but a full plasma.
The plasma is assumed to consist of thermalized visible sector particles;
the hidden sector must be (nearly) unpopulated. Otherwise, the Goldstino
couples directly to those particles and will thermalize with that sector,
leading to the usual gravitino problem.
This is clearly a condition %\foot{We are not aware of a reference to
%this fact in the literature on supersymmetric inflation models.}
 on inflation models: the inflaton, whose decay
is responsible for reheating the universe, must not couple to the hidden
sector (except through unavoidable gravitational couplings). This is
sufficient to ensure that the decay of the inflaton will thermalize
the visible sector fields, but not the hidden sector.

In such a situation, we wish to study the approach of the Goldstino to
thermal equilibrium.
The idea that we will use here is to attempt to compute the inclusive
production rate,\foot{We thank T. Banks for discussions on this point.}
where we
sum over all final states including a Goldstino\foot{States with more than one
Goldstino will be parametrically suppressed.} and average over initial states
by using a thermal density matrix $\rho_\beta=e^{-\beta H}/Z$. We
thereby avoid having to interpret individual scattering amplitudes.

The Goldstino couples to the visible sector through the supercurrent,
with
\eqn\superc{{1\over F}\partial_\mu S^\mu_{\alpha,vis}
= -{i\over M} {\cal O} \equiv -{i\over M}(aW'_A(\phi^*)\psi^A_\alpha+
b G_{\mu\nu}(\sigma^{\mu\nu}\lambda)_\alpha+\ldots)}
where the coefficients $a,b$ are of order one.
The total transition probability per unit 4-volume
for production of single gravitinoes may be written:
\eqn\transprob
{\eqalign{w_\chi=&{1\over M^2}\int {d^3 q\over (2\pi)^3 2E_q}
\sum_{spins}  u(\vec q,s)_\alpha\bar u(\vec q,s)_\eta
\int d^4 x\; e^{iq\cdot x}
\Tr\left[\rho_\beta\; {\cal O}_\alpha (x)  \bar{\cal O}_\eta (0)\right]\cr
=&{1\over M^2}\int {d^3 q\over (2\pi)^3 2E_q}
\int d^4 x\; e^{iq\cdot x}\; \tr\; \qslsh\;\langle
{\cal O}(x) \bar {\cal O}(0)\rangle_\beta .\cr}}
The use of the r.h.s. of eq. \superc\ is justified here, since we are
bracketing on on-shell states, which satisfy the full equations of
motion.
In the first line of \transprob,
we have written the result in terms of Dirac spinors
and $q^\mu$ is the Goldstino 4-momentum. Since we are interested
in $T\gg m_{3/2}$ we have taken $q^2=0$. We note that the thermal
matrix element $\langle{\cal O}(x) \bar{\cal O}(0)\rangle_\beta\equiv\Tr
\left[\rho_\beta {\cal O}(x) \bar{\cal O}(0)\right]$ is
a function of real times, but is {\it not} a time-ordered quantity.
Thus we can also write this expression in terms of the discontinuity
(or the imaginary part) of the gravitino self energy\Weldon\
\eqn\disc{w_{\chi}={1\over M^2}\int {d^3 q\over (2\pi)^3 2E_q}
{1\over e^{\beta\omega}+1} \tr\; \qslsh\;
2 {\rm Im}\;\langle {\cal O} \bar {\cal
O}\rangle_{q,\beta}\equiv {\tilde w_\chi\over M^2}
.}

We are thus lead to an expression for the total production rate which is
 suppressed by ${1/M^2}$ as long as the matrix element is
non-singular in the limit $M\to \infty$, \ie\ as long as there is no
resonant mixing. In what follows we will argue
that this is indeed the case. Notice that a behaviour $w_\chi\sim (1/M)^r$
with $0<r<2$ would be already physically interesting; the result of
Ref. \fisch\ corresponds to $r=0$.

In the limit $M\to\infty$, the (thermalized) visible
sector, in which $\langle {\cal O}\bar{\cal O}\rangle_\beta$ is calculated,
corresponds to an exactly supersymmetric theory: the MSSM without soft
supersymmetry breaking terms. The singularity required to modify the
$1/M^2$ behaviour of eq. \disc, would be physically interpreted
as a mode in a fermionic channel ${\cal O}$, which propagates undamped
on the light-cone. The only modes which propagate undamped at finite
$T$, according
to prime principles, are Goldstones corresponding to spontaneously
broken symmetries. In fact these have to be undamped at
$(\omega,k)=(0,0)$, in order to saturate the broken Ward identities.
For $(\omega,k)\not =(0,0)$ we have no physical reason to expect an
undamped mode in an interacting theory, and in fact experience shows
that these modes are always damped.
It is known that finite temperature leads to a
spontaneously broken representation of supersymmetry and the
appearance of a Goldstino-like mode. Indeed fermionic Green's
functions like $\langle {\cal O} \bar{\cal O}\rangle_\beta$ are singular at
$(\omega,k)=(0,0)$. Fortunately it is possible to characterize in a
quite general way the properties of this singularity.
 As shown in the next section, its dispersion relation in a weakly
coupled theory at high $T$ is given by $\omega/k=1/3$, \ie\ strictly
{\it off} the light-cone. In other words, in the limit in which the
Goldstino is undamped, only the real part of $\langle {\cal OO}\rangle
$ (and not the imaginary one) has a $1/k$ singularity when approaching
 $(\omega, k)=(0,0)$ from the light-cone. (Indeed, cfr. (3.3), a much stronger
singularity would be needed to affect $\tilde w_\chi$).
 We will thus conclude that no resonant mixing can arise
with the zero-temperature Goldstino in eq. \disc, so that
$w_\chi\sim 1/M^2$. In the Appendix we discuss an example where the
Goldstone singularity gets arbitrarily close to the light-cone as
$k\to 0$, so that a deviation from the naive behaviour for a total
emission rate does indeed take place.

Indeed, since $\tilde w_\chi$ is well defined as
$M\to \infty$ it will also be calculable in perturbation theory at
weak (gauge and Yukawa) coupling. The leading contributions come
from two-loop diagrams involving A-terms ({\it cf.} the first term in
eq. \superc), as well as contributions from the gaugino mass
(second term in eq. \superc). One-loop diagrams
do not contribute to $\tilde w_\chi$ in the susy limit: they
correspond to two body decay processes with goldstino emission, which
are forbidden by phase space. The differential rate
${\rm d}^3w_\chi/{\rm d}^3 k$,
when computed at momenta $<gT$ requires some care to be taken.
The matrix element \disc\
will in general contain infrared divergences when calculated
in perturbation theory. These divergences signal
a breakdown in naive perturbation theory when the energy carried
by a given line (or by all lines entering a vertex)
in a graph is of order $g_s T$; fortunately, it is known
how to resum the perturbation series.\hardthermloops\ The result, for nonzero
external momenta $(\omega,q)$ is that all modes are screened at order
$g_s T$, including gauge particles. This is consistent with
our general argument that $\tilde w_\chi$ is finite. We stress once more
that in order to avoid the $1/M^2$ suppression for Goldstino
production, $\tilde w_\chi$ {\it (even after
resummation) } would have to be {\it infinite}
(we have taken $M\rightarrow\infty$). There is one
remaining subtlety in the case of non-Abelian gauge theory; namely
in the integration region $\omega,q\lsim g_s^2 T$, we expect further
infrared divergences. These are similar to those
encountered in the static limit, where for example, the calculation of
the free energy of hot QCD irrevocably breaks down at some
order.\foot{For recent progress on this problem, see Ref. \eric.}
Thus ultimately, we cannot calculate the partial rate in this regime.
If something dramatic were to happen to the rate in this momentum region,
 it would happen  in a perturbatively incalculable way. But
again, consistent with the general physical argument discussed above,
 we feel that this is implausible. For instance, it is
known that in hot QCD\leptoprod\ the quasiparticle
spectrum in the quark propagator can lead to enhancements in the
partial rate at low energies. However, it seems reasonable that higher
order effects lead to a finite width for these modes which would
cut off any zero-energy singularity and, if we may use the analogy
here, $\tilde w_\chi$ would then indeed be finite.

\newsec{Temperature Goldstinoes}

In this final section, we will discuss some aspects of the
behaviour of supersymmetric theories at finite temperature and
the appearance of thermal Goldstino modes.
Supersymmetry is broken at finite temperature because of the non-zero
energy density in the ground state; the breaking thus appears spontaneous
and there exists a fermionic Goldstone mode.
In this section we study the physics of this mode;
we begin by considering theories in which supersymmetry is unbroken
at zero temperature. Much of what we have to say is a review of portions
of the literature, although we believe some of our comments are new.

In Ref. \girard, it is argued that temperature treats fermions and bosons
differently, so supersymmetry is expected to
be broken. They suggest that it is explicitly broken by the boundary
conditions (the antiperiodic boundary condition for fermions leads to
a problem in imaginary time Ward identities: the Grassmann variational
parameter $\epsilon$ cannot be taken to be a constant) and that no
massless fermion mode need arise.

In Refs. {\boyan,\aoyboy}, it was realized that in order to study
dynamical effects, one must consider the problem in the real-time
finite temperature formalism.
It was established that the Ward identities at finite
temperature are essentially those of the zero temperature theory
with matrix elements interpreted as finite temperature correlation
functions. In particular, Boyanovsky\boyan\ established a Ward identity
involving a supercurrent-fermion matrix element and found a fermionic
pole with no mass gap when $\langle F\rangle_\beta$ is non-zero.

Let us consider the problem now in more general terms.
The superalgebra relation
\eqn\qqdagger{\left\{ Q_\alpha,Q_{\dot\alpha}^\dagger\right\}\,=\,
2\sigma_{\alpha,\dot\alpha}^\mu P_\mu}
implies that a state is invariant under supersymmetry if and only
if its energy is zero. Then, when the ground state itself has non-zero
energy, supersymmetry is spontaneously broken and a Goldstone
fermion is present in the spectrum. The existence of the Goldstino
is established by writing eq. \qqdagger\ in its local form, as
a Ward identity
\eqn\ssdagger {\partial_\mu\langle T\left ( S^\mu_\alpha(x)
{S^\nu_{\dot\alpha}}^\dagger(0)\right )\rangle\,=\,
\delta^4(x)\langle \theta^\nu_\rho(0)\rangle \sigma^\rho _{\alpha
\dot\alpha}}
where $S^\mu$ and $\theta^{\mu\nu}$ are respectively the supercurrent
and the stress energy tensor. By integrating eq. \ssdagger,  a
non-zero righthand side implies the existence of a zero momentum
singularity in the current-current correlator. Moreover, Lorentz
invariance of the vacuum constrains the singularity to be of the form
$1/\parslsh$, signaling the presence of a massless fermion.
Notice, though, that eq.
\ssdagger, is a local operator identity determined by the dynamics
of the system, and, as such, is valid on any stationary, translation
invariant state.
In particular in a thermal bath,
$\langle \theta\rangle\not = 0$, and we expect long range correlations
in the system, and  the absence of a mass gap. The absence of Lorentz
invariance, though, does not
allow one to establish the nature of the singularity in a model
independent way \strocchi. For instance in the free massless
Wess-Zumino model, the left hand side in eq. \ssdagger\ is easily
evaluated and the singular part has the same form as the
Klimov-Weldon \weldon\ self-energy from hard thermal loops. The form is
involved, with logarithmic singularities on the light-cone, but the
only ``pole-type'' singularity is at the single point
$\omega=0$, $k=0$ (where $\omega$
and $k$ are respectively energy and three-momentum). Thus
there is long-range order, but without a propagating wave.

In fact, as also noticed in Ref. \aoyama, eq. \ssdagger\ is the
supersymmetric analogue of the Ward identity expressing the
spontaneous breakdown of Lorentz invariance in a thermal bath
\eqn\sound{\int d^4x\;\partial_x^\lambda\langle T(M_{\lambda\mu\nu} (x)
\theta_{\rho\eta}(y))\rangle_\beta
=\langle g_{\mu\rho}\theta_{\nu\eta}(y)-g_{\nu\rho}\theta_{\mu\eta}(y)
+ g_{\mu\eta}\theta_{\rho\nu}(y)-g_{\nu\eta}\theta_{\rho\mu}(y)
\rangle_\beta}
where $M_{\mu\nu\rho}$ is the Lorentz current.
In a thermal bath, $\langle \theta_{\rho\eta}\rangle \not\propto
g_{\rho\eta}$ and
the right-hand side is non-zero. Then the physical
interpretation of Eq. \sound\ is that there are propagating
hydrodynamic fluctuations (sound waves) corresponding to variations
in the stress-energy density.
These modes correspond to local boosts of the thermal bath, which in the
limit of infinite wave-length cost zero energy.\foot{Notice the
subtlety in counting Goldstone modes for this space-time symmetry;
only one of three modes propagates.} In addition we also
know from hydrodynamics that these modes correspond to propagating
waves only at wave-lengths that are much bigger than the mean free
path (see for example ref. \weinberg). In particular, in a free field
theory, the mean free path is infinite, and eq. \sound\ does not give
rise to any interesting physical wave at finite $(\omega, k)$. This
is consistent with our findings for eq. \ssdagger. By analogy, it
also suggests that in an interacting supersymmetric theory a
hydrodynamic fermionic wave will describe the fluctuations
of the system above some long, but finite, wavelength. It also
suggests, that, by a reasoning that parallels the one in
hydrodynamics, one should be able to derive the wave equation
for the Temperature Goldstino in a quite general way.

Let us
briefly recall the procedure in hydrodynamics. One considers a fluid
that in stationary conditions in its restframe
is described by a stress energy tensor
$\theta_{\mu\nu}=pg_{\mu\nu}+(p+\rho)g_{0\mu}g_{0\nu}$,
where $p$ and $\rho$ are respectively the pressure and energy density.
The Goldstone mode of broken Lorentz (or Galilean) symmetry
corresponds to small local boosts
of the quiescent system which are described by a
space-time dependent velocity three-vector $v_i$. To lowest order in
$v$, the energy momentum in the ``global'' rest frame of the bath
will have the boosted form
\eqn\boost{{\tilde\theta}_{\mu\nu}=
\theta_{\mu\nu}+\Delta_v\left [\theta_{\mu\nu}\right ]=
pg_{\mu\nu}+(p+\rho)
(g_{0\mu}g_{0\nu}-v_ig_{i\mu}g_{0\nu}-v_ig_{0\mu}g_{i\nu})}
The continuity equation for $\tilde\theta$ together with the equation
of state of the fluid, then give the equation of propagation
of the Goldstone mode,\foot{Notice that to derive the equation
the derivatives of $p,\rho$ have to be considered non zero and of order
$v$, \ie\ the longitudinal part of the Goldstone field $v$ ``mixes''
with other scalar modes. This is because when Lorentz invariance is
broken, there can be mixing terms between the Goldstone field
and heavier modes which contain just one and not two time derivatives.
For example, in the liquid helium case, the Lagrangian contains
the term $\mu \Phi\partial_t \pi$ where $\mu$ is the chemical
potential, $\Phi$ the real part of the condensate and $\pi$ the
Goldstone phase. Integrating out the massive $\Phi$ produces a
relevant quadratic piece for $\pi$. As we will see this phenomenon
does not take place for Goldstinoes due to the linearity in
$(\partial_t,\partial_x)$ of the wave equation.} the sound velocity
equaling $v_s=\sqrt{\partial p/\partial\rho}$.\foot{Here we are
considering zero chemical potential, so that sound waves
are associated with fluctuations in the energy density.}

Let us now consider the case of supersymmetry.
In a thermal bath the density of supercurrent $\langle
S_\alpha^\mu\rangle$ vanishes at equilibrium. In the presence of a
space time dependent Goldstino oscillation $\xi$ the supercurrent
$\bar S$ is  given by
\eqn\wave{\bar S_{\dot\alpha}^\mu=\xi^\alpha (x)\left \{
Q_\alpha,\bar S_{\dot\alpha}^\mu\right \}= \theta^{\mu\nu}
(\bar \sigma_\nu\xi(x))_{\dot\alpha}}
and conservation of the current corresponds to the Goldstino
wave equation. We thus obtain the dispersion relation
\eqn\disper{\omega=\left ( {p\over \rho}\right )k.}
The velocity of the Goldstino $v_G=p/\rho$ should be compared to the
velocity of sound $v_s=\sqrt{\partial p/\partial \rho}$. In a weakly
coupled theory
we have respectively in the non-relativistic and relativistic regimes:
$v_G=v_s^2=T/m$ and $v_G=v_s^2=1/3$. In the intermediate regimes,
though, the relation $v_G=v_s^2$ does not hold.

Eq. \disper\ agrees with the explicit calculation of Ref. \gund,
where the fermion self-energy in the
Wess-Zumino model was studied at low temperature. However
we suspect that in the treatment in Ref. \gund, and
also {\boyan,\aoyboy},  higher order corrections are not completely
under control. Though these effects will not change
eq. \disper\ (apart from the form of higher order corrections to the vacuum
energy and pressure), we suspect that they can significantly affect
the residue of the Goldstino pole. In what follows we briefly motivate
our doubts. For the sake of the argument we will stick to the low
temperature case which was studied in Ref. {\gund,\boyan,\aoyboy}. Eq.
\ssdagger\ suggests that the order parameter for
supersymmetry breaking is a composite operator, \ie\ the stress-energy
tensor. We thus expect the Goldstino to lie predominantly in a
composite channel, suggesting that the description of this mode
in terms of the elementary fields in the lagrangian is indeed
non-perturbative. In Refs. \boyan,\aoyboy,
it was noticed that in the massive Wess-Zumino model, $\langle\phi\rangle$ is
displaced from its zero temperature value $\phi_0=m/g$
and we have a vacuum expectation value for the auxiliary field
$\langle F\rangle=\partial_\phi W\not = 0$ generated at finite temperature;
thus the Goldstino has a non-zero overlap with the elementary
fermion field in the theory.
The fermion and boson masses, as calculated from the tree lagrangian,
are thus split by $\Delta=m_A^2-m_\psi^2=m_\psi^2-m_B^2=gF$, where $A$, $B$
and $\psi$ are respectively the scalar, pseudoscalar and fermion.
The 1-loop effective fermion self energy $\Sigma(\omega,k)$
is consistently calculated by using
the free propagators with the masses corrected by the
background $F$-field value. At zero momentum in real time one finds
$\Sigma\sim g^2/\Delta$.
In our case, we must have $m+ \Sigma(0,0)=0$
as required by the broken-supersymmetry Ward identity.\boyan\
The dominant contributions to $\Sigma(\omega,k)$ for $(\omega, k)\ll
( \Delta/T,\Delta/\sqrt{mT})$, are determined by denominators of the form
\eqn\denom {{1\over{E_\psi^2-(E_A-\omega)^2}}\sim {1\over{\Delta
+2pk-2E_A\omega}}}
where $p$ is the integration 3-momentum.  The above denominator
is the same as that of hard thermal loops, apart from the ``infrared
regulator'' $\Delta$.
It is however clear that any higher order
contribution affecting the fermion and boson propagators can affect
the result, since it is determined in lowest order by the ``small''
$\Delta$. For instance, the genuine 1-loop corrections to the scalar
masses at low temperature
(\ie\ not those coming from $\langle F\rangle\not=0$) are
not smaller than order $\Delta$. Of course the fact that individual
corrections affect the result does not mean that the full correction will,
especially since SUSY cancellations might be expected. Nonetheless
it is not clear to us how to implement the ``nominally'' higher order
effects in a manifestly SUSY consistent manner. This is especially
clear at high temperature where the 1-loop corrections to propagators
dominate the tree level. When one is interested in calculating
processes at low momenta $\omega\sim gT$ the resummation of hard
thermal loops \pisarski\ is the correct approach. However, for ultralow
momenta $\omega\ll g^2T$ we are not aware of any developed technique.

The damping rate of this mode was also calculated in Ref. \gund\
and was found to be exponentially
small at small momentum\foot{Of course we know that the damping rate
must be exactly zero at $\omega,k=0$,
since there must be a Goldstone singularity.}
 $\gamma_D\sim exp(-\beta\Delta/2k)$, where $\Delta$
is the effective fermion-boson mass splitting induced by temperature.
This exponential suppression arises presumably only at one loop.
At this order the attenuation of the Goldstino is determined
by a three body process, in which a boson (fermion) in the bath
absorbs a thermal Goldstino and becomes a fermion (boson). Since fermions
and boson effective masses have a splitting $\Delta$, energy-momentum
conservation requires an energy $O(\Delta/\omega)$ for the absorbing
particle, thus the Boltzmann suppression at low $\omega$.
At higher orders the thermal Goldstino can be absorbed in a
four- (or higher) body process. In this case, we do not expect the above phase
space exponential suppression, and presumably the only suppression
at low $(\omega,k)$ will be due to the fact that the Goldstino is derivatively
coupled.
% \ie\ $\gamma_D\sim \omega^2T/\Delta$. We thus expect that above
%a typical wavelength $\sim \Delta/T$, the damping is negligible and
%the wave is indeed physically relevant.

The dispersion relation was obtained in Ref. \gund\ by expanding
around $(\omega,k)=0$, and is apparently not sensitive to $\Delta$.
Fortunately, for the discussion of gravitino production,
we need only have control of
the dispersion relation \disper. The fact that the ``singularity''
lies far off the light cone is clear evidence that it cannot play
any relevant role in gravitino production, as the mode will not
mix with the gravitino. Given this, the behaviour of the Goldstino
damping $\gamma_D$ cannot play any role in our discussion. Nonetheless
it would be physically interesting to find a way to calculate this
quantity.

Our analysis should be contrasted with the case studied in
 the Appendix, where the mode that leads to enhanced axion production
is given by the pion. It that case, indeed, the relevant mode is not a
purely collective one but
a thermally ``dressed'' massless particle. Thus, not surprisingly,
the mass shell is very close to the light-cone and
a partial enhancement takes place.

We would like to thank T. Banks, M. Dine, G. Farrar, R. Pisarski,
S. Raby, and S. Thomas for useful
discussions. We are indebted to the Aspen Center for Physics where
this work was begun.

\newsec{Appendix}

In this Appendix we show the details of the pion-axion mixing example
which was mentioned in the text.
We focus on the case of two light quark flavors $u,d$.

The coupling between the axion field $a$ and hadronic matter is given,
at lowest order, by the $(a, \pi_0)$ mass matrix
\eqn\matrice
{{\cal L_{a\pi}}  = (a,\,\pi_0) \left(\matrix{m_{aa}^2 & m_{a\pi}^2\cr
 m_{a\pi}^2 & m_{\pi\pi}^2\cr}\right )\left(\matrix {a\cr\pi_0\cr}\right)=
 B(a,\,\pi_0) \left(\matrix{{m_u+m_d\over f_a^2}&{m_u-m_d\over f_af_\pi}\cr
 {m_u-m_d\over f_af_\pi}&{m_u+m_d\over f_\pi^2}\cr}\right )
\left(\matrix {a\cr\pi_0\cr}\right)}
where $f_a\gg f_\pi$ are the axion and pion decay constants, while
$B=<\bar q q>\sim \Lambda_{\rm QCD}^3$. We are interested in the rate
for axion production from a thermal bath of hadrons. To lowest order
in $m_{a\pi}^2$ and using the formalism of Section 3, the rate of
axion production  per unit time and volume is
\eqn\rate{
{{\rm d}n_a\over {\rm d} t}=m_{a\pi}^4\int {{\rm d}^3k_a\over
{(2\pi)^3 2 E_a}} {1\over \exp{(E_a/T)}-1}i {\rm Disc} \bigr
[G_{\pi\pi}(E_a,k_a)\bigl ]}
Where
 $G_{\pi\pi}$ is the pion propagator in the thermal bath. This function has
been calculated at finite temperature up to ${\cal O}(1/f_\pi^4)$
in the chiral expansion.\schenk\ Even in the chiral limit, both the real
and imaginary parts of $G_{\pi\pi}^{-1}(\omega,k)$ are non-vanishing on the
light-cone. The only surviving singularity is at $(\omega,k)=(0,0)$,
consistent with Goldstone's theorem. The thermal self-energy
$\Pi_{\pi\pi}(\omega,k)$ can be written as
\eqn\self
{\Pi_{\pi\pi}(\omega,k)= \bigl ( R+iI\bigr ) {T^4\over f_\pi^4}k^2+
{T^2\over f_\pi^2}{\cal O}(K^2,m_\pi^2)+i{T^4\over f_\pi^4}
{\cal O}(K^2,m_\pi^2)}
where $K^2=\omega ^2-k^2$, while $R$ and $I$ are numerical
coefficients  and $R>0$.
In order to
illustrate the infrared enhancement effect we consider the limit $m_\pi\ll
T\ll f_\pi$. In this regime the dominant contribution comes from
relativistic pions. Then, as will become clear below, we
need only consider the term proportional to $R+iI$ in eq.\self. The
other terms are subleading with respect to the tree-level chirality breaking
terms in $G_{\pi\pi}$. The
discontinuity of the pion propagator then reads
\eqn\discont{{i\rm Disc}G_{\pi\pi}(E_a,k_a)={2f_\pi^4\over T^4}{Ik_a^2\over
\bigl [(m_a^2-m_\pi^2){f_\pi^4\over T^4}+Rk_a^2\bigr ]^2+I^2k_a^4}}
The above expression behaves like $1/k_a^2$ in the range $k_a\gg
k_{\rm inf}\sim m_\pi (f_\pi^2/T^2{\sqrt R})$. Then, for $k_a$ in this range,
the integral in eq. \rate\ is linearly infrared divergent, related to
the bosonic nature of the pion, and its proximity to the light-cone. This
divergence is cut-off at momenta of order $k_{\rm inf}$.
Neglecting numerical factors, the total rate behaves like
\eqn\nonan{{{\rm d}n\over {\rm d}t}\propto m_{a\pi}^2{f_\pi^2\over
m_\pi T}\propto m^{3\over 2}}
displaying an enhanced and non-analytic behaviour in the quark mass $m$.
Notice also the growth of eq. \nonan\ at low temperature.
As temperature is lowered the pion density gets smaller, but the
overlap of a pion excitation with a massless particle improves due
to the smaller real and imaginary parts in $\Pi_{\pi\pi}$. Then there
is coherence on a longer scale and a pion is more likely to oscillate
into an axion.

Of course, the above results can be qualitatively
interpreted by considering the quantum mechanics of pion-axion
oscillations taking into account damping phenomena. For a pion emitted
at $t=0$, the probability of finding an axion at time $t$ is given by
$P(t)=\epsilon^2\sin^2(\Delta m^2 t/2E)$, where $\epsilon$ is the mixing
angle and $\epsilon \Delta m^2=m_{a\pi}^2$. However, in the presence
of damping, this will be limited by the finite coherence length of
the pions. For a damping rate $\Gamma\gg m^2/E$,
we have $P_{\pi\to a}\sim m_{a\pi}^4/(E\Gamma)^2$.
Then we simply find the rate of axion production
\eqn\qual{{d n_{a}\over {\rm d}t{\rm d}^3k}\sim {m_{a\pi}^4\over
E^2\Gamma^2} \Gamma  n_{\pi}, }
qualitatively consistent with the above more formal analysis.

\if\norefs n\listrefs\fi
\end